\documentclass[doublecol]{epl2} 

\usepackage{amsmath} 
\usepackage{amssymb}
\usepackage{mathrsfs}
\usepackage{graphicx}
\usepackage{epstopdf}  

\title{Spontaneous rotating vortex rings in a parametrically driven\\ polariton fluid}

\author{J.~O.~Hamp \inst{1,2}\thanks{ E-mail: \email{joh28@cam.ac.uk}} \and A.~K.~Balin \inst{1,3}\thanks{E-mail: \email{andrew.balin@physics.ox.ac.uk}} \and F.~M.~Marchetti \inst{4} \and D.~Sanvitto \inst{5} \and M.~H.~Szyma\'nska \inst{1,6}\thanks{E-mail: \email{m.szymanska@ucl.ac.uk}}}

\shortauthor{J.~O.~Hamp \etal}

\institute{                    
  \inst{1} Department of Physics, University of Warwick, Coventry,
  CV4 7AL, UK \\
	\inst{2} TCM Group, Cavendish Laboratory, University of Cambridge, Cambridge, CB3 0HE, UK \\
	\inst{3} Rudolf Peierls Centre for Theoretical Physics, University of Oxford, Oxford, OX1 3NP, UK \\
  \inst{4} Departamento de F\'isica Te\'orica de la Materia
  Condensada \& Condensed Matter Physics Center (IFIMAC), Universidad
  Aut\'onoma de Madrid, Madrid 28049, Spain \\
  \inst{5} NNL, 
	Istituto Nanoscienze-CNR, 
	Via Arnesano, 
	73100 Lecce, Italy \\
	\inst{6} Department of Physics and Astronomy, University College
  London, Gower Street, London, WC1E 6BT, UK  \\
}


\abstract{
We present the theoretical prediction of spontaneous rotating
  vortex rings in a parametrically driven quantum fluid of
  polaritons -- coherent superpositions of coupled quantum well
  excitons and microcavity photons. These rings arise
  not only in the
absence of any rotating drive, but
  also in the absence of a trapping potential, in a model known to map
  quantitatively to experiments. We begin by proposing a novel
  parametric pumping scheme for polaritons, with circular
  symmetry and radial currents, and characterize the resulting nonequilibrium
  condensate. We show that the system is unstable to spontaneous
	breaking of circular symmetry via a modulational instability, following which a vortex ring with large
	net angular momentum emerges, rotating in one of two topologically
  distinct states. Such rings are robust and carry distinctive
  experimental signatures, and so they could
  find applications in the new generation of polaritonic devices.}

\begin{document}

\maketitle

\section{Introduction}

Macroscopically coherent quantum fluids exhibit excitations in the
form of quantized vortices -- topological defects in the order
parameter describing the condensed phase. The ubiquity of quantized
vortices has become increasingly apparent since their prediction in
superfluids some decades ago \cite{onsager1949nuovo}, and they are now
known to play a key role in the physics of equilibrium systems such as
ultracold atomic gases, liquid helium, and type-II superconductors
\cite{leggett2006quantum}. 

More recently, condensation of bosonic
quasiparticles, such as microcavity polaritons
\cite{kasprzak2006bose}, has been observed. Polaritons are coherent
superpositions of coupled quantum well excitons and
microcavity photons. They have a finite lifetime, so
their condensation is intrinsically nonequilibrium: continuous
repopulation from an external source is necessary to balance photonic
losses from the cavity. In the optical parametric oscillator (OPO)
regime \cite{stevenson2000continuous, baumberg2000parametric},
polaritons are resonantly injected into a pump state by a coherent laser field,
before undergoing parametric scattering to signal and idler states. The sum of the phases of the signal and idler is locked by the pump, but their relative phase is otherwise free, and any explicit choice by the system breaks U(1) gauge symmetry. In this sense, though out-of-equilibrium, the system can be thought of as a quantum condensate in the same way as an equilibrium Bose-Einstein condensate -- both are characterized by the appearance of a Goldstone mode \cite{wouters2007goldstone}.

Parametrically driven polariton fluids have been shown to have nonequilibrium
superfluid properties \cite{amo2009collective, amo2009superfluidity}, and exhibit
complex and varied behavior such as quantized vortices and
persistent currents \cite{sanvitto2010persistent,
  marchetti2010spontaneous, tosi2011onset}. 
	Under certain external perturbations, quantized vortices can form nontrivial patterned structures: for example, in ultracold atomic gases under a rotating drive,
or type-II superconductors in an external magnetic field
\cite{leggett2006quantum}. However, the appearance of vortices, and their subsequent pattern formation, is not expected to occur
spontaneously, without an external injection of angular momentum of some kind.

In this letter, we present the theoretical prediction of spontaneous vortex
rings in a nonequilibrium polariton fluid driven in the parametric regime. 
These vortex
rings have nonzero net angular momentum, and appear due to the strong driving and dissipation in the system -- not only in the
absence of any rotating drive, but also in the absence
of a trapping potential.

We begin by proposing a novel and experimentally viable parametric
pumping scheme for polaritons, with circular symmetry and radial
currents. We then characterize the resulting circularly-symmetric
condensate. In many cases the system is unstable to spontaneous breaking of circular symmetry via a modulational instability, and undergoes the formation of vortex rings in the
absence of any rotating drive. These rings have nonzero net angular momentum and rotate 
in one of two degenerate but topologically distinct states of opposite chirality.
Their presence leads to side bands in the photoluminescence spectrum
of the system \cite{borgh2010spatial}, which constitutes an accessible
experimental signature of such an ring, in addition to possible
direct imaging with time-resolved techniques.

These are the first examples of spontaneous vortex rings in
parametrically driven polariton fluids, and find their natural place
in the wider framework of pattern formation in out-of-equilibrium
systems \cite{cross1993pattern}. They are fundamentally different from the spontaneous vortex-antivortex rings recently observed in other studies of polaritons under different pumping schemes \cite{tosi2012geometrically, manni2013spontaneous, hivet2014interaction}, in that here all vortices have the same vorticity, and correspondingly the rings carry nonzero net angular momentum. Spontaneous vortex arrays with nonzero net angular momentum had been predicted to
arise \cite{keeling2008spontaneous} in a simplified model of an
incoherently-pumped polariton condensate, but there they necessitated
an harmonic trapping potential. Our rings, beyond being the first
examples of spontaneous vortex rings of this type in a parametrically driven
polariton fluid, provide great encouragement for the first
experimental observation in any polariton system, since they carry
distinctive experimental signatures, and dispense with the need for
harmonic trapping potentials~\cite{balili2007bose}. Moreover,
they occur in a model which is known to map quantitatively to
experimental systems, unlike those of incoherently-pumped systems,
where the roles of the excitonic reservoir and thermalization are not
well understood. Due to their controllable, topologically robust nature, with large angular momentum, they 
have the potential to find applications in the new generation of polaritonic devices \cite{de2012control, franchetti2012exploiting, ballarini2013all}.

\section{Methods} 

Semiconductor microcavity polaritons in the parametrically driven (OPO) regime are known to be
well described by a generalized Gross-Pitaevskii (GP)
equation~\cite{whittaker2005numerical} that is derivable as the
mean-field limit of a microscopic theory where exciton-exciton and
exciton-photon interactions, and the pumping and decay processes, are
all treated explicitly. The details are reviewed in
Ref.~\cite{carusotto2013quantum} and references therein. 

The coupled
exciton ($X$) and cavity photon ($C$) fields $\psi_{X,C} (\mathbf{r},t)$
decay with rates $\kappa_{X,C}$ respectively, and mix with strength
$\Omega_R/2$, where $\Omega_R$ is the Rabi splitting between the upper
(UP) and lower (LP) branches of the polariton dispersion at zero
detuning. The excitonic dispersion $\omega_X$ is taken to be 
constant $\omega_X^0$, and the photonic dispersion is given by
$\omega_C=\omega_C^{0} -\frac{\nabla^2}{2m_C}$, where $m_C$ is the
cavity photon mass. The generalized GP equation is then (we set
$\hbar=1$ throughout):
\begin{multline}
  i\partial_t \begin{pmatrix} \psi_X \\ \psi_C \end{pmatrix}
  = \begin{pmatrix} 0 \\ F_p \end{pmatrix}
\\ + 
  \begin{pmatrix} \omega_X -i \kappa_X + g_X|\psi_X|^2& \Omega_R/2 \\
  \Omega_R/2 & \omega_C -i \kappa_C \end{pmatrix} \begin{pmatrix}
    \psi_X \\ \psi_C
  \end{pmatrix},
\label{eq:model}
\end{multline}
where $g_X$ is the strength of the exciton-exciton interaction and
$F_p(\mathbf{r}, t)$ is the pump field.  The traditional OPO system is
driven by a coherent continuous-wave pump, $F_{p}(\mathbf{r},t) =
\mathcal{F}_{p} (r) \, e^{i (\mathbf{k}_{p} \cdot \mathbf{r} - \omega_{p}
  t)}$, with $\mathbf{k}_p$ fixed in one particular direction (e.g.~the
$x$-direction), and a smoothed top-hat or Gaussian spatial profile
$\mathcal{F}_{p} (r)$ of strength $f_{p}$ and full width at
half-maximum (FWHM) $\sigma_{p}$. The fields $\psi_{X,C}$ and pump
strength $f_{p}$ can be rescaled by $\sqrt{\Omega_R/(2 g_X)}$ so that
the exciton interaction strength $g_X$ is unity. In the present work
we use $m_C=2 \times 10^{-5} m_e$, $\Omega_R = 4.4$ meV (both typical
of GaAs-based microcavities), and zero detuning.  

Equation
\eqref{eq:model} describes a system that exhibits rich and complex
phenomena arising from the subtle interplay between pumping, decay,
nonlinearity, and the system being finite size. It is well known that
for finite size pump profiles, no approximate analytic analysis that
captures the OPO physics -- including the instability to spontaneous
vortex formation -- is possible \cite{whittaker2005numerical,
  whittaker2007vortices, marchetti2010spontaneous}. We proceed by
solving Eq.~\eqref{eq:model} via a fifth-order adaptive-step
Runge-Kutta algorithm for a system of size $120 \times 120 \;
\mu\mathrm{m}$ discretized onto $2^8 \times 2^8$ points in
space. The pump is chosen to have wave vector $|\mathbf{k}_p | =1.6 \;
\mu\mathrm{m}^{-1}$ and energy $\omega_p = - 0.44$ meV, in resonance
with the point of inflection of the LP dispersion branch after
blueshifting \cite{whittaker2005effects} of the polariton dispersion.


\section{Pumping scheme}

\begin{figure}
\includegraphics[width=1\linewidth]{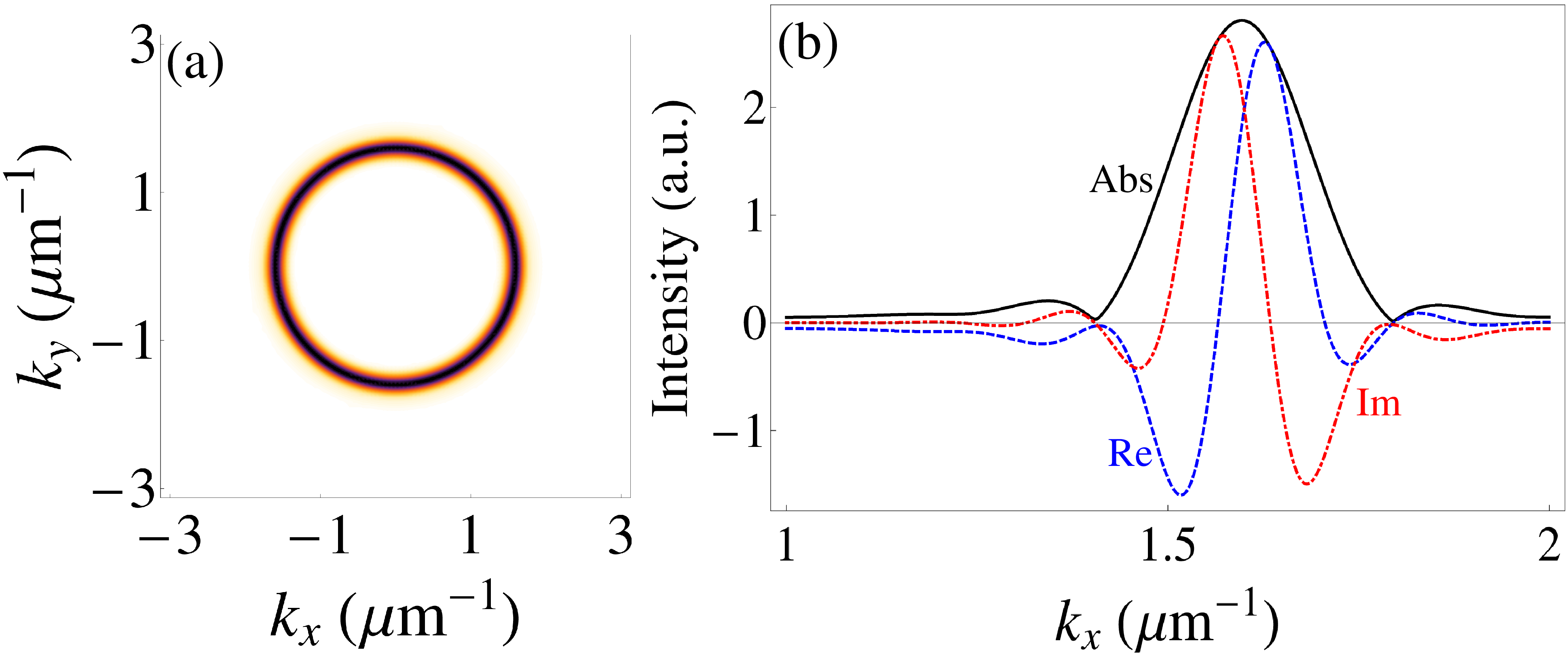}
\caption{(Colour on-line) External pump used to drive the system in the OPO regime. In
  real space the pump is given by $F_{p}(\mathbf{r},t) = \mathcal{F}_{p}
  (r) \, e^{i (k_p r - \omega_{p} t)}$, with a smoothed top-hat
  profile $\mathcal{F}_{p} (r)$, i.e., constant radial currents of
  strength $|\mathbf{k}_p|$. (a) The absolute value of the Fourier
  transform of the pump field, in two-dimensional momentum space. The
  pump constitutes a ring of radius $|\mathbf{k}_p|$. (b) A cut along
  $k_y=0$, showing the details of the ring profile in momentum space:
  absolute value, and real and imaginary parts.}
\label{fig:pump}
\end{figure}

We build upon previous theoretical work in polariton fluids
\cite{marchetti2012vortices} by proposing a new parametric pumping
scheme $F_p(\mathbf{r},t)$. We use $F_p(\mathbf{r},t)=\mathcal{F}_{p} (r)
\, e^{i(k_p r - \omega_p t)}$, and a smoothed top-hat spatial profile
$\mathcal{F}_{p} (r)$. This pumping configuration corresponds to a
circularly symmetric profile, with constant radial currents given by
$|\textbf{k}_p| \hat{\textbf{r}}$. The absolute value of the Fourier
transform (zeroth order Hankel transform) of the pump field in
two-dimensional momentum space $|F_p(\mathbf{k})|$ is shown in
Fig.~\ref{fig:pump} (a); it is a ring of radius $|\mathbf{k}_p|$. The
profile of the ring is shown in detail in a cut along $k_y=0$ in
Fig.~\ref{fig:pump} (b), along with the real and imaginary parts.
This pumping scheme could be realized in experiments with the aid of
an axicon lens, or engineered using the phase of the laser pump, and a
spatial light modulator. It leads to the generation of an OPO state
that is circularly symmetric both in real and in momentum
space.

However, in order to investigate the dynamical stability of the
steady-state symmetric system, a small
perturbation has to be added. We use a weak probe
field in resonance with the signal mode, but any other weak
symmetry-breaking perturbation would work equally
well. 
In what follows we only consider the
photonic component of the polariton field since it is the quantity
measured in experiments.

\section{Results}

\begin{figure}
\includegraphics[width=1\linewidth]{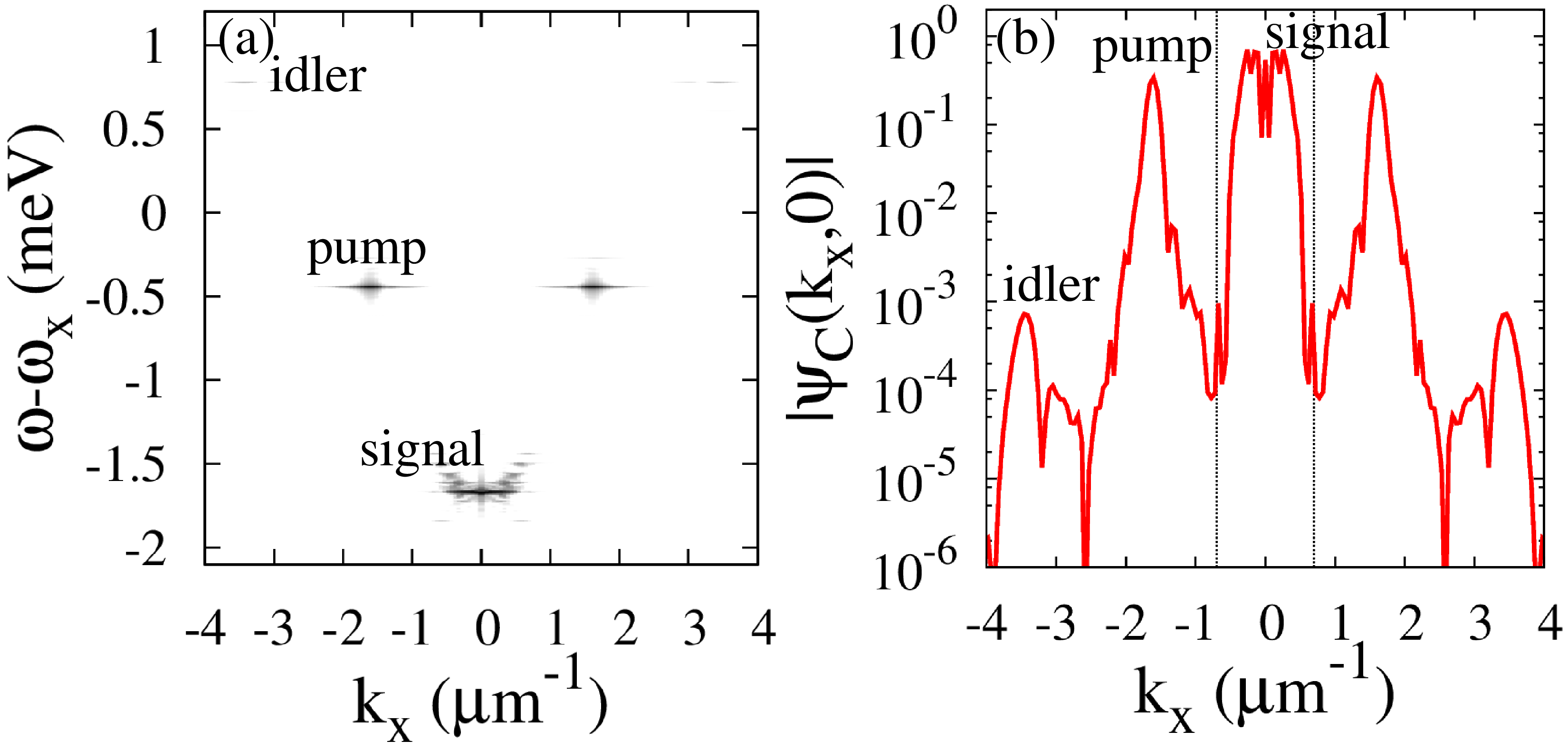}
\caption{(Colour on-line) (a) The full photoluminescence spectrum of the symmetric OPO system with $\sigma_p=35 \,
  \mu\mathrm{m}$, along $k_y=0$. Polaritons are injected coherently into the pump state, and scatter parametrically into the signal and idler states.  The intensity scale is logarithmic. (b) The momentum population of the symmetric system,
  $|\psi_C(\mathbf{k})|$, along $k_y=0$. The strong occupations of the pump, signal, and
  idler modes as created by the OPO process can be clearly
  seen. Each can be obtained individually by filtering in a suitable window in momentum. The dashed lines indicate the window in which we filter to obtain the signal mode.}
\label{fig:opo}
\end{figure}

The full photoluminesence spectrum of the symmetric system with pump
spot size $\sigma_p=35 \, \mu\mathrm{m}$ is shown in
Fig.~\ref{fig:opo} (a) (along $k_y=0$). The generation of OPO is
evidenced by the presence of signal and idler states to which
polaritons injected at the pump state parametrically scatter, the
signal lying at low momentum and energy, and the idler at high
momentum and energy \footnote{The idler state is weak, and especially
  so in our system since it lies on a ring of large radius in
  two-dimensional momentum space, and is therefore correspondingly
  diluted in any one-dimensional cut.}.  The total photonic emission
in momentum space $|\psi_C(\mathbf{k})|$ of the same system can be seen
in Fig.~\ref{fig:opo} (b), as a cut along $k_y=0$. Again, the
macroscopic populations of the pump (at $|\mathbf{k}_p|=1.6 \, \mu
\mathrm{m}^{-1}$), signal (at $ |\mathbf{k}_s| \sim 0$) and idler (at
$|\mathbf{k}_i| \sim 2|\mathbf{k}_p|$) states are visible. We are
interested in the behaviour of the signal mode population, which we
obtain by filtering in energy or momentum, as is done in
experiments. The idler has a conjugate behaviour to the signal (i.e.,
if vortices appear in the signal, antivortices will appear in the
idler). The dashed lines in Fig.~\ref{fig:opo} (b) represent the
window of $\pm 0.7 \, \mu \mathrm{m}^{-1}$ around zero momentum which
we use to filter out the the signal mode and obtain the signal field
in real space $|\psi_C^s(\mathbf{r}, t)| \, e^{i \phi_C^s(\mathbf{r},
  t)}$, where $\phi_C^s(\mathbf{r}, t)$ is the phase of the signal wave
function.

The parametric scattering process occurs (`switches on') for certain ranges of detunings
and pump momenta \cite{whittaker2005effects}, only above some
threshold pump strength $f_p^{\mathrm{th}}$.  We find the signal mode
population to appear at $f_p \equiv f_p^{\mathrm{th}}$ and increase
above threshold up to some maximum value, before beginning to
disappear (`switching off') at around $f_p = 1.6 f_p^\mathrm{th}$. The
maximum total signal intensity occurs for a pump strength of around
$f_p = 1.5 f_p^{\mathrm{th}}$. However, the shape of the signal in
real space generally becomes more complicated and less uniform on
increasing $f_p$ too far above threshold, and after symmetry breaking,
the time evolution is not stable. The regime slightly above threshold
is where interesting behavior has been observed in previous work
\cite{marchetti2010spontaneous}, and it is also the regime in which we
observe steady-state vortex ring solutions. We observe vortex rings
for $1.1 \leq f_p/f_p^\mathrm{th}\leq 1.25$, and in the following we
use $f_p=1.2 f_p^\mathrm{th}$ unless stated otherwise. 

\begin{figure}
\includegraphics[width=1\linewidth]{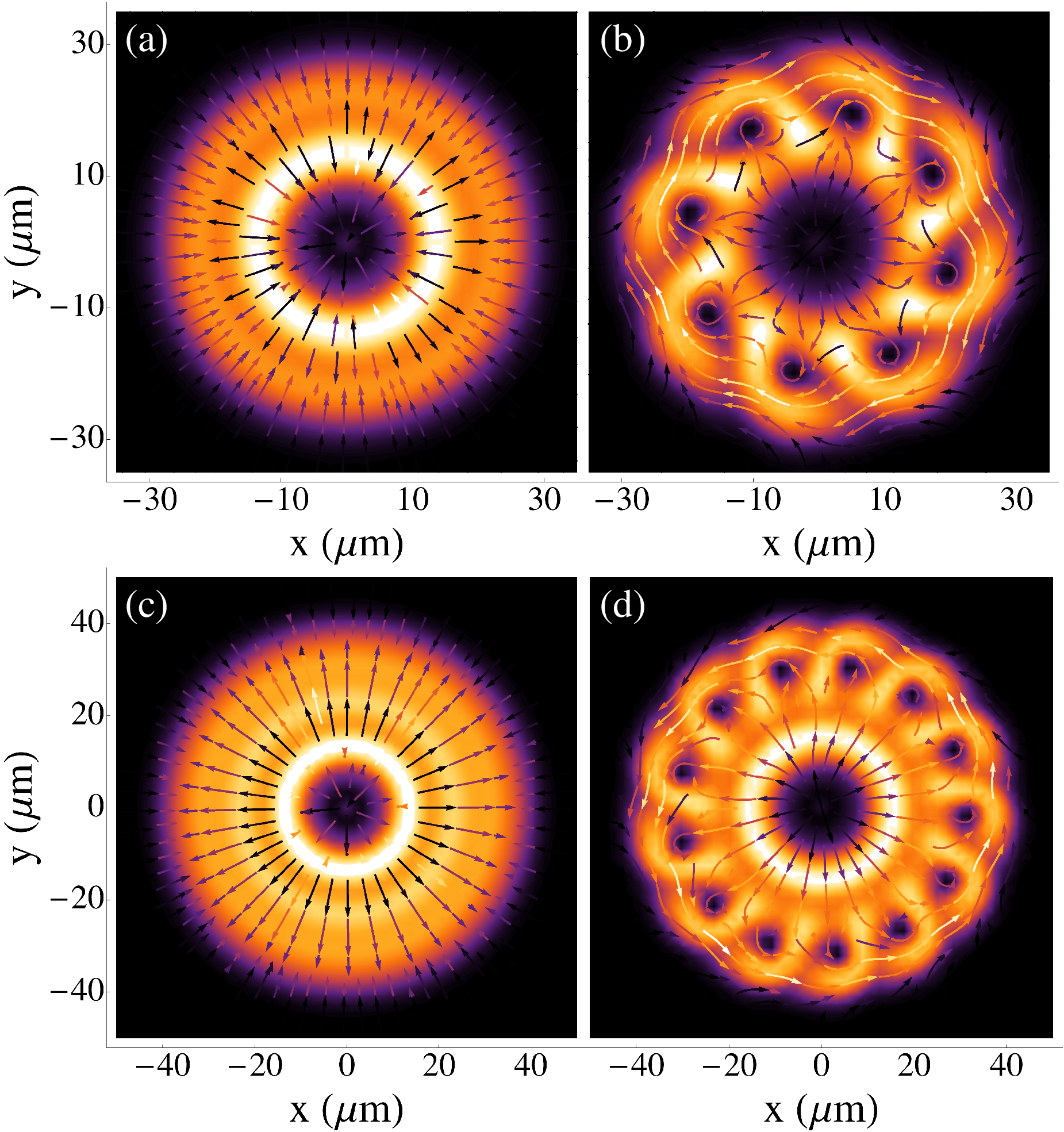}
\caption{(Colour on-line) The photonic component of the signal mode population of the
  system in real space $|\psi_C^s(\mathbf{r})|$. (a), (b): With
  $\sigma_p=35 \, \mu\mathrm{m}$. (a) In the absence of a vortex ring
  (before symmetry breaking). (b) In the presence of an 8-(anti)vortex
  ring. (c), (d): With $\sigma_p=46 \, \mu\mathrm{m}$. (c) In the
  absence of a vortex ring (before symmetry breaking). (d) In the
  presence of a 13-vortex ring. Superimposed are the supercurrents
  $\mathbf{j}(\mathbf{r})$. In the symmetric cases, the dominant signal
  current at small nonzero momentum has been subtracted to reveal the
  more complex underlying currents.}
\label{fig:currents}
\end{figure}

The photonic
signal emission in real space $|\psi_C^s(\mathbf{r})|$ of the symmetric
system with $\sigma_p=35 \, \mu\mathrm{m}$ and $\sigma_p=46 \,
\mu\mathrm{m}$, can be seen in Figs.~\ref{fig:currents} (a) and (c),
respectively. Superimposed are the supercurrents $\mathbf{j}(\mathbf{r})
\equiv |\psi^s_C|^{2} \, \nabla \phi^s_C(\mathbf{r})$, which result
from the interplay of spatially non-uniform pumping and decay. The
small but dominant (radial) signal currents at nonzero $k=0.25 \;
\mu\mathrm{m}^{-1}$ have been subtracted to reveal the more complex
underlying steady-state current structure which moves particles from
gain- to loss-dominated regions. There are sharp changes in the density profile of the condensate, and at the border of higher-and lower-density regions, discontinuities in the direction of the underlying 
currents. The pump state population is uniform and homogeneous throughout, as directly imposed by the top-hat spatial profile of the pump field, with strong currents directed radially outwards. It is not significantly modified by the presence of vortices in the signal and idler.

Once the system has evolved to a steady-state regime with balanced
pumping and decay, we examine dynamical stability by adding a small perturbation. For this we add a
Gaussian probe $F_{p}(\mathbf{r}, t) \rightarrow F_{p}(\mathbf{r}, t) +
F_{pb}(\mathbf{r}, t)$, resonant with the signal mode but of much
smaller dimensions ($\sigma_{pb} \sim 3 \; \mu\mathrm{m}$), $10^{-3}$
times weaker than the pump, positioned such that $\mathbf{r} \times \mathbf{k}_{pb}=0$, so as not to impose a
preferred direction of symmetry breaking. The strength of the perturbation influences the transitory regime but
has no bearing on the final solution. The discontinuities in the underlying current direction in the symmetric system correspond to polaritons travelling with smaller or larger velocities relative to the average. 
 The presence of such radial
counter-propagating currents, due to the strong driving, decay, and nonlinearity, means that system is dynamically unstable. This can be understood as similar to the situation in other nonlinear systems, such as water waves approaching a shore, where the dispersion of the waves eventually leads to an instability. 

Following the perturbation, vortices enter the condensate and stabilize there in
an ordered ring.
The
ring can be formed of either vortices or antivortices and rotates
clockwise or anticlockwise (respectively). 
However, neither rotation direction is preferred energetically -- thus, the explicit choice made by the system spontaneously breaks circular symmetry.
The physical mechanism of vortex ring formation is fundamentally spontaneous, however, in experiments, symmetry would be broken explicitly, e.g., by any small in-plane asymmetry of the cavity field. 
In the simulations, the direction of rotation can be controlled if desired, by breaking the symmetry explicitly in a particular direction, e.g., by imparting some
momentum in that directions. 
However, if a noise perturbation is used that does not explicitly prefer any rotation direction, then for the same parameters but different noise realizations, the system will rotate sometimes left, sometimes right.
In the presence of the vortex ring, the unstable radial counter-propagating currents are eliminated, and the flow of the condensate is controlled by the vortex ring. These vortex ring
solutions are dynamically stable to noise or additional small
perturbations, and we simulate the dynamics for very long times, $\sim
15 \; \mathrm{ns}$.

An example of a steady-state vortex ring can be seen in
Fig.~\ref{fig:currents} (b), with the supercurrents
$\mathbf{j}(\mathbf{r})$ superimposed. The system is the same as that in
Fig.~\ref{fig:currents} (a). The condensate contains 8 antivortices
and rotates anticlockwise as a rigid body with angular frequency $\Omega =
9.91$ GHz. In the
presence of the vortex ring, taking into account the currents in the
whole signal state, there are no longer discontinuous changes in the
current direction throughout the condensate as in the symmetric case
(see Fig.~\ref{fig:currents}, (a); (c)), and the vortex ring
dominates the currents in the system. This 8-vortex ring solution is
robust to varying $\sigma_p$ and $f_p$ by up to around $\pm 5$
\%. When increasing (decreasing) $\sigma_p$ by more than this, rings
with greater (fewer) numbers of vortices can be generated, such as the
13-vortex ring in Fig.~\ref{fig:currents} (d), using a 30 \% larger
pump spot size (the same system as in Fig.~\ref{fig:currents}
(c)). This is due to the fact that vortices are organized such that their separation is set by the
healing length, which is approximately constant, so that a larger condensate
can support more vortices. Outside the window of $1.1 \leq
f_p/f_p^\mathrm{th}\leq 1.25$, however, the symmetry-broken system is
unstable: vortices drift in and out of the condensate without ordering
on the longest timescales we can simulate for. This is analogous to
the lack of a steady-state for other effects in previous simulations
of the OPO system \cite{marchetti2010spontaneous}. However, if the
pump is strong ($f_p \gtrsim1.5$ over a range of $\sigma_p$ values),
and the signal population comparatively uniform, then the circularly
symmetric solution is dynamically stable and no symmetry breaking
occurs; this corresponds to when the condensate is at its most dense.

\begin{figure}
\includegraphics[width=1\linewidth]{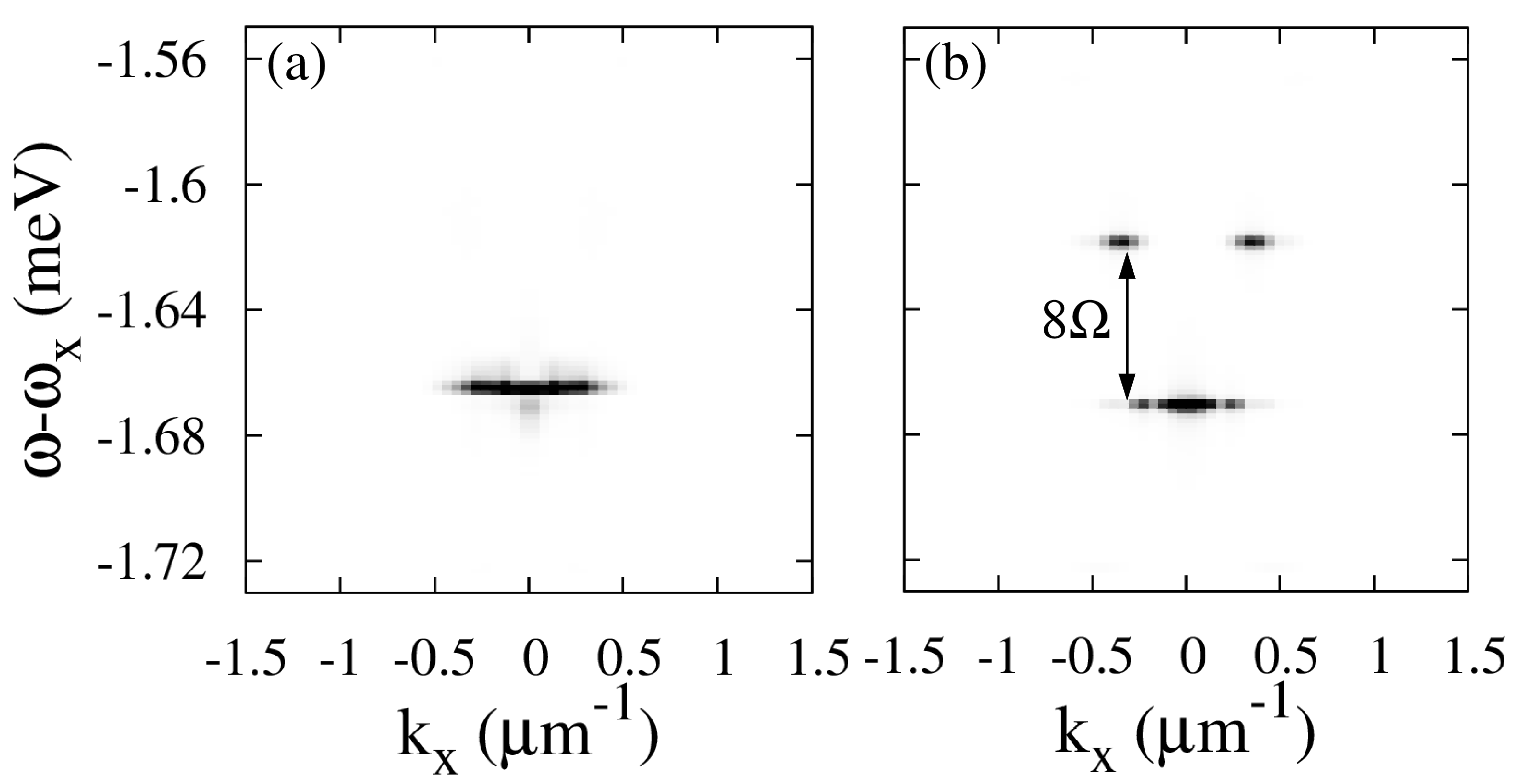}
\caption{(Colour on-line) The photoluminescence of the system close to the signal mode with
  $\sigma_p=35 \, \mu\mathrm{m}$, along $k_y=0$. (a) In the absence of
  an ring (before symmetry breaking). (b) In the presence of an
  8-(anti)vortex ring. In both cases the signal mode population
  around $k=0$ can be seen. In the presence of the ring two
  additional side-bands, arising from rotation of the condensate,
  appear $\Delta E = 8 \Omega$ above the signal energy.}
\label{fig:spectra}
\end{figure}

The appearance of the vortex ring can be understood as being due to a
modulational instability of the symmetric system: in general, pumping and decay causes the frequencies of a stationary, homogeneous state to acquire an imaginary part, leading to growth of the associated high angular momentum modes, and vortex nucleation \cite{keeling2008spontaneous,borgh2012robustness}. As such, our rings are predicted
\cite{borgh2010spatial} to give rise to side bands in the
photoluminescence spectrum, shifted in energy away from the condensate
mode by $\Delta E = n \Omega$, where $n$ is the number of vortices in
the ring. The photoluminescence spectrum near the signal (along
$k_y=0$) of the system corresponding to Fig.~\ref{fig:currents} (a) is
shown in Fig.~\ref{fig:spectra} (a). The signal population can be seen
around $k=0$, $\omega- \omega_X=-1.67$ meV. The spectrum of the same
system in the presence of an 8-vortex ring (Fig.~\ref{fig:opo} (b))
is shown in Fig.~\ref{fig:spectra} (b). Two side bands lying above the
signal state can be clearly seen.  The energy shift $\Delta E$ of the
side bands relative to the signal state is found to be $0.051 \pm
0.002$ meV, in very good agreement with the theoretical value of $8
\Omega = 0.052$ meV. We find similarly good agreement in the case of
rings with different numbers of vortices. For example, for a
12-vortex ring solution, we find the frequency of rotation to be $\Omega =
8.10$ GHz, giving a theoretical energy shift of $12 \Omega = 0.064$
meV, and we observe the side bands to be shifted by $0.064\pm 0.002$
meV away from the signal state. Based on our simulations, we believe
that the vortex ring could be directly imaged, and its rotation
measured, using state-of-the-art time-resolved techniques. For
example, for a 8-vortex ring with $\sigma_p=35 \, \mu\mathrm{m}$ we
predict a period of rotation of 634 ps, whilst for a 12-vortex ring
with $\sigma_p=43.5 \, \mu\mathrm{m}$ we predict 776 ps, which should
be long enough to allow direct imaging. However, for experiments
which rely upon time-integrated measurements of photonic emission,
direct observation of rotating vortex rings may be more
difficult. The presence of side bands in the photoluminescence
spectrum provides an alternative experimental signature of such an
ring. Another possible method of directly imaging a rotating vortex
ring in experiments using time-integrated measurements is via the
defocused homodyne imaging scheme discussed in
Ref.~\cite{borgh2012robustness}, which exploits the presence of the
side bands in the spectrum.

\section{Conclusion}

We have presented the theoretical predicition of spontaneous rotating vortex ring
formation in a parametrically driven polariton fluid. This has been
achieved by proposing a novel pumping scheme with circular symmetry
and radial currents. The pumping scheme results in a circularly
symmetric condensate with steady-state currents arising from the
interplay of spatially nonuniform pumping and decay. We find that
the system is dynamically unstable to spontaneous symmetry breaking
and undergoes the formation of rotating vortex rings in the absence
of any rotating drive or fields. The rings have large net angular momentum and can rotate in either direction. 
Side bands in the photoluminescence
spectrum of the system constitute an experimental signature of such an
ring alternative to direct imaging with time-resolved techniques. Due 
this and the fact that they dispense with the need for harmonic trapping 
potentials, they provide great encouragement 
for the first experimental observation of spontaneous vortex rings in 
any polariton system. Since the rotation states can be controlled, 
and are topologically distinct, the system has the potential to find
applications in polaritonic devices as a controllable two-state
system, or for storing large, definite 
values of angular momentum that can be used to direct polariton flow.

\acknowledgments
We thank J.~Keeling for helpful discussions. JOH and AKB both acknowledge support from University of Warwick Scholarships.  FMM
  acknowledges financial support from the programs MINECO
  (MAT2011-22997), and CAM (S-2009/ESP-1503). DS acknowledges the 
  project ERC Polaflow. MHS acknowledges support from EPSRC (EP/I028900/1 and EP/K003623/1).

\bibliography{paper_mod}

\end{document}